\documentclass[landscape,a4paper]{article} 
\usepackage{multicol}
\usepackage[english]{babel}
\usepackage[utf8]{inputenc}
\usepackage{amsmath}
\usepackage{amssymb}
\usepackage{graphicx}
\usepackage[landscape,margin=2.5cm]{geometry}
\usepackage{tabularx}
\usepackage{longtable}
\usepackage{rotating}
\usepackage{listings}
\usepackage[usenames,dvipsnames]{xcolor}
\usepackage{hyperref}

\newsavebox\ltmcbox

\hypersetup
{
    pdfauthor={Tobias Knopp and Martin M\"oddel},
    pdfsubject={Magnetic Particle Imaging Data Storage Format},
    pdftitle={MDF: Magnetic Particle Imaging Data Format},
    pdfkeywords={MDF, MPI, HDF5},
    pdfproducer={Section for Biomedical Imaging Hamburg University of Technology, Section for Biomedical Imaging University Medical Center Hamburg-Eppendorf},
    pdfcreator={Section for Biomedical Imaging Hamburg University of Technology, Section for Biomedical Imaging University Medical Center Hamburg-Eppendorf}
}

\lstdefinelanguage{MDF}%
  {morekeywords={%
    Bool,Int64,Float64,Number,String%
      },%
   sensitive=true,%
}[keywords,comments,strings]%

\newcommand{\inl}[1]{\lstinline[columns=fixed]{#1}}
\newcommand{\inltab}[1]{{\ttfamily\bfseries\color{blue}#1}}
\newcommand{\inlvar}[1]{{\ttfamily#1}}
\newcommand{\adj}{{\vdash \hspace*{-1.752mm} \dashv}}

\begin{document}

\title{MDF: Magnetic Particle Imaging Data Format}
\newcommand{\version}{2.1.0}

\author{
T.~Knopp$^{1,2}$, T.~Viereck$^3$, G.~Bringout$^4$, M.~Ahlborg$^5$, A.~von~Gladiss$^5$, C.~Kaethner$^5$, A.~Neumann$^5$, P.~Vogel$^6$, J.~Rahmer$^7$, M.~M\"oddel$^{1,2}$ \\ \\
$^1$Section for Biomedical Imaging, University Medical Center Hamburg-Eppendorf, Germany\\
$^2$Institute for Biomedical Imaging, Hamburg University of Technology, Germany\\
$^3$Institute of Electrical Measurement and Fundamental Electrical Engineering, TU Braunschweig, Germany\\
$^4$Physikalisch-Technische Bundesanstalt, Berlin, Germany\\
$^5$Institute of Medical Engineering, University of  Lübeck, Germany\\
$^6$Department of Experimental Physics 5 (Biophysics), University of Würzburg, Germany\\
$^7$Philips GmbH Innovative Technologies, Research Laboratories, Hamburg, Germany
}

\maketitle
\begin{center}
Version \textbf{\version}
\end{center}

\begin{abstract}
Magnetic particle imaging (MPI) is a tomographic method for determining the spatio-temporal distribution of magnetic nanoparticles. In this document, a file format for the standardized storage of MPI and magnetic particle spectroscopy (MPS) data is introduced. The aim of the Magnetic Particle Imaging Data Format (MDF) is to provide a coherent way of exchanging MPI and MPS data acquired with different devices worldwide. The focus of the MDF is on sequence parameters, measurement data, calibration data, and reconstruction data. The format is based on the hierarchical document format in version 5 (HDF5). This document discusses the MDF version \version, which is not backward compatible with version 1.x.y.
\end{abstract}

\begin{multicols}{2}		

\section{Introduction} \label{Sec:Introduction}

The purpose of this document is to introduce a file format for exchanging Magnetic Particle Imaging (MPI) and Magnetic Particle Spectroscopy (MPS) data. The Magnetic Particle Imaging Data Format (MDF) is based on the hierarchical document format (HDF) in version 5 \cite{hdf5}. HDF5 is able to store multiple datasets within a single file providing a powerful and flexible data container. To allow an easy exchange of MPI data, one has to specify a naming scheme within HDF5 files which is the purpose of this document. In order to create and access HDF5 data, an Open Source C library is available that provides dynamic access from most programming languages. MATLAB supports HDF5 by its functions \inlvar{h5read} and \inlvar{h5write}. For Python, the \inlvar{h5py} package exists, and the Julia programming language provides access to HDF5 files via the \inlvar{HDF5} package. For languages based on the .NET framework, the \inlvar{HDF5DotNet} library is available.

The MDF is mainly focused on storing measurement data, calibration data, or reconstruction data together with the corresponding sequence parameters and metadata. Even though it is possible to combine measurement data and reconstruction data into a single file, it is recommended to use a single file for each of the following dataset types:
\begin{enumerate}
\setlength{\itemsep}{0pt}
\item Measurement data
\item Calibration data
\item Reconstruction data
\end{enumerate}

\subsection{Datatypes \& Storage Order}

MPI parameters are stored as regular \textit{HDF5 datasets}. \textit{HDF5 attributes} are not used in the current specification of the MDF. For most datasets, a fixed datatype is used, for example the drive-field amplitudes are stored as \inl{H5T_NATIVE_DOUBLE} values. Whenever a data type has a big- and little-endian version, the little-endian data type should be used. For our convenience, we refer to the HDF5 datatypes \inl{H5T_STRING}, \inl{H5T_NATIVE_DOUBLE}, and \inl{H5T_NATIVE_INT64} as \inltab{String}, \inltab{Float64} and \inltab{Int64}. Boolean data is stored as \inl{H5T_NATIVE_INT8}, which we refer to as \inltab{Int8}. 

The datatype of the MPI measurement and calibration data offers more freedom and is denoted by \inltab{Number}, which can be any of the following HDF5 data types: \inl{H5T_NATIVE_FLOAT}, \mbox{\inl{H5T_NATIVE_DOUBLE},} \mbox{\inl{H5T_NATIVE_INT8},} \inl{H5T_NATIVE_INT16}, \inl{H5T_NATIVE_INT32}, \inl{H5T_NATIVE_INT64} or a complex number as defined below. In the case that the calibration data is compressed, the indices are stored as \inltab{Integer}, which can be any of the following HDF5 data types: \mbox{\inl{H5T_NATIVE_INT8},} \inl{H5T_NATIVE_INT16}, \inl{H5T_NATIVE_INT32}, \inl{H5T_NATIVE_INT64}.

Since storing complex data in HDF5 is not standardized, we define a compound datatype \inl{H5T_COMPOUND} in HDF5 with fields \inl{r} and \inl{i} using one of the above mentioned number types to represent the real and the imaginary part of a complex number. This representation was chosen because it is also the default behavior for complex numbers in Python using numpy and h5py as well as in Julia using the HDF5.jl package. \inltab{Complex128} refers to the complex compound type with base type \inl{H5T_NATIVE_DOUBLE}.

For later identification of a data set, we store three Universally Unique Identifiers (UUIDs) (RFC~4122)~\cite{leach2005universally} in its canonical textual representation as 32 hexadecimal digits, displayed in five groups separated by hyphens \inlvar{8-4-4-4-12} as for example \inlvar{ee94cb6d-febf-47d9-bec9-e3afa59bfaf8}. For the generation of the UUIDs, we recommend to use version 4 of the UUID specification.

Whenever multidimensional data is stored, dimensions are arranged in a way that cache is utilized best for fast reconstruction or fast frame selection for example. The leading dimension in the MDF specification is slowest to access and the last dimension is the fastest to access (contiguous memory access). That also means that dependent on the memory layout of your programming language the order of the dimensions will be just as in the MDF specifications (row major) or reversed (column major). Please take this into account when reading or writing MDF files, which includes the usage of HDF5 viewers.

\subsection{Units}

With one exception, physical quantities are given in SI units. The magnetic field strength is reported in T$\mu_0^{-1} = 4 \pi$Am$^{-1}\mu_0^{-1}$. This convention has been proposed in the first MPI publication \cite{Gleich2005} and consistently been used in most MPI related publications. The aim of this convention is to report the numbers on a Tesla scale, which most readers with a background in magnetic resonance imaging are familiar with.

\subsection{Optional Parameters}

The MDF has 8 main groups in the root directory and 2 sub-groups. We distinguish between optional and non-optional groups as well as optional, non-optional, and conditional parameters. 

Any optional parameter can be omitted, whereas any non-optional parameter in a non-optional group is mandatory. Conditional parameters are linked to Boolean parameters and have to be provided, if these parameters are true and can be omitted if they are false. If a parameter is optional, non-optional, or conditional is indicated by yes, no, or the corresponding Boolean parameter.

If a group is optional all of its parameters may be omitted, if this group is not used. The groups \inlvar{/}, \inlvar{/study}, \inlvar{/experiment}, \inlvar{/scanner}, \inlvar{/acquisition} contain mostly metadata and are mandatory. The \inlvar{/tracer} group is only mandatory if magnetic material has been placed in the MPI system. The groups \inlvar{/measurement}, \inlvar{/calibration}, and \inlvar{/reconstruction} are all optional. In case of calibration measurements, the \inlvar{/calibration} group is mandatory. The reconstruction data is stored in \inlvar{/reconstruction}. 

\subsection{User defined parameters}

It is possible to store user defined parameters in an MDF. For instance one may want to store the temperature of the room in which your MPI device is operated. In this case, one is free to add new parameters to any of the existing groups. Moreover, if necessary, one can also introduce new groups. In order to be able to distinguish these datasets and groups from the specified ones, it is mandatory to use the prefix ``\inlvar{\_}'' for all parameters and groups. As an example, one could add a new group \inlvar{/\_room} that includes the dataset \inlvar{\_temperature}. Using the prefix ``\inlvar{\_}'' will ensure that the stored dataset is compatible with future versions of the MDF.

\subsection{Naming Convention}

Several parameters within the MDF are linked in dimensionality. We use short variable names to indicate these connections. The following table describes the meaning of each variable name used in this specification.
\setbox\ltmcbox\vbox{
	\makeatletter\col@number\@ne
	\noindent \begin{longtable}{p{0.12\columnwidth}p{0.8\columnwidth}} 
		\textbf{Variable} & \textbf{Number of}  \\ \hline
		$A$ &  tracer materials/injections for multi-color MPI \\ \hline
		$N$ &  acquired frames ($N=O+E$), same as a spatial position for calibration\\ \hline
		$E$ &  acquired background frames ($E=N-O$)\\ \hline
		$O$ &  acquired foreground frames ($O=N-E$)\\ \hline
		$B$ &  coefficients stored after sparsity transformation ($B\leq O$)\\ \hline
		$J$ &  periods within one frame\\ \hline
		$Y$ &  partitions of each patch position\\ \hline
		$C$ &  receive channels \\ \hline
		$D$ &  drive-field channels \\ \hline
		$F$ &  frequencies describing the drive-field waveform \\ \hline
		$V$ &  points sampled at receiver during one drive-field period\\ \hline
		$W$ &  sampling points containing processed data (\mbox{$W=V$} if no frequency selection or bandwidth reduction has been applied)\\ \hline
		$K$ &  frequencies describing the processed data ($K=V/2+1$ if no frequency selection or bandwidth reduction has been applied)\\ \hline
		$Q$ &  frames in the reconstructed MPI data set\\ \hline
		$P$ &  voxels in the reconstructed MPI data set\\ \hline
		$S$ &  channels in the reconstructed MPI data set\\ \hline
	\end{longtable}
	\unskip
	\unpenalty
	\unpenalty}
\unvbox\ltmcbox

\subsection{Contact}

If you find mistakes in this document or the specified file format or if you want to discuss extensions or improvements to this specification, please open an issue on GitHub:\\
\hspace*{1cm}\url{https://github.com/MagneticParticleImaging/MDF}\\
As the file format is versionized it will be possible to extend it for future needs of MPI. The current version discussed in this document is version \version.

\subsection{arXiv}
As of version 1.0.1, the most recent release of these specifications can also be also found at:\\
\hspace*{1cm}\url{https://arxiv.org/abs/1602.06072}\\
If you use MDF, please cite this document using the arXiv reference, which is also available for download as \texttt{MDF.bib} from GitHub.

\subsection{Code examples}		 
  		  
If you want to get a basic impression of how to handle MDF files you can visit the gitub repository of the MDF project:\\		
  \hspace*{1cm}\url{https://github.com/MagneticParticleImaging/MDF}\\	
There you will find the example directory, which contains a code example written in Julia, MATLAB, and Python. More details can be found in the \texttt{README} of the repository.
 
\subsection{Reference Implementation}		 

A reference implementation for a high level MDF access is available at:\\
  \hspace*{1cm}\url{https://github.com/MagneticParticleImaging/MPIFiles.jl}\\	
\inlvar{MPIFiles.jl} \cite{knopp2019mpifiles} is a package written in the programming language Julia \cite{Bezanson2012,Bezanson2014,Bezanson2014a}. It can read MDF V1, MDF V2, and the dataformat of Bruker MPI systems using a common interface. It also provides functions to convert MDF V1 or Bruker MPI files into MDF V2.

\end{multicols}

\clearpage

\section{Data (group: \inlvar{/})}

\begin{multicols}{2}
\paragraph{Remarks:} Within the root group, the metadata about the file itself is stored. Within several subgroups, the metadata about the experimental setting, the MPI tracer, and the MPI scanner can be provided. The actual data is stored in dedicated groups about measurement data and reconstruction data.
\end{multicols}

\setlength\extrarowheight{5pt}
\noindent \begin{tabularx}{\columnwidth}{lllllX} 
\textbf{Parameter} & \textbf{Type} & \textbf{Dim} & \textbf{Unit/Format} & \textbf{Optional} & \textbf{Description} \\ \hline 
\inlvar{time} & \inltab{String} & 1 & yyyy-mm-ddThh:mm:ss.ms & no & UTC creation time of MDF data set \\ \hline
\inlvar{uuid} & \inltab{String} & 1 & 3170fdf8-f8e1-4cbf-ac73-41520b41f6ee & no & Universally Unique Identifier (RFC 4122) of MDF file \\ \hline 
\inlvar{version} & \inltab{String} & 1 & \version & no & Version of the file format \\ \hline
\end{tabularx}
\setlength\extrarowheight{0pt}

\subsection{Study Description (group: \inlvar{/study/}, non-optional)}

\begin{multicols}{2}
	\paragraph{Remarks:} A study is supposed to group a series of experiments to support, refute, or validate a hypothesis. The study group should contain \inlvar{name}, \inlvar{number}, \inlvar{uuid}, and \inlvar{description} of the study.
\end{multicols}

\setlength\extrarowheight{5pt}
\noindent \begin{tabularx}{\columnwidth}{lllllX} 
\textbf{Parameter} & \textbf{Type} & \textbf{Dim} & \textbf{Unit/Format} & \textbf{Optional} & \textbf{Description} \\ \hline 
\inlvar{description} & \inltab{String} & 1 & & no & Short description of the study \\ \hline
\inlvar{name} & \inltab{String} & 1 & & no & Name of the study \\ \hline
\inlvar{number} & \inltab{Int64} & 1 & & no & Number of the study\\ \hline
\inlvar{time} & \inltab{String} & 1 & yyyy-mm-ddThh:mm:ss.ms & yes & UTC creation time of study \\ \hline
\inlvar{uuid} & \inltab{String} & 1 & 295258fe-b650-4e5f-96db-b83f11089a6c & no & Universally Unique Identifier (RFC 4122) of study \\ \hline 
\end{tabularx}
\setlength\extrarowheight{0pt}

\subsection{Experiment Description (group: \inlvar{/experiment/}, non-optional)}

\begin{multicols}{2}
\paragraph{Remarks:} For each experiment within a study a \inlvar{name}, \inlvar{number}, \inlvar{uuid}, and \inlvar{description} have to be provided. Additionally, the name of the \inlvar{subject} imaged and the flag \inlvar{isSimulation} indicating if data has been obtained via simulations have to be stored.
\end{multicols}

\setlength\extrarowheight{5pt}
\noindent \begin{tabularx}{\columnwidth}{lllllX} 
\textbf{Parameter} & \textbf{Type} & \textbf{Dim} & \textbf{Unit/Format} & \textbf{Optional} & \textbf{Description} \\ \hline 
\inlvar{description} & \inltab{String} & 1 & & no & Short description of the experiment \\ \hline
\inlvar{isSimulation} & \inltab{Int8} & 1 & & no & Flag indicating if the data in this file is simulated rather than measured \\ \hline
\inlvar{name} & \inltab{String} & 1 & & no & Experiment name \\ \hline
\inlvar{number} & \inltab{Int64} & 1 & & no & Experiment number within study\\ \hline
\inlvar{subject} & \inltab{String} & 1 & & no & Name of the subject that was imaged \\ \hline 
\inlvar{uuid} & \inltab{String} & 1 & f96dbc48-1ebd-44c7-b04d-1b45da054693 & no & Universally Unique Identifier (RFC 4122) of experiment \\ \hline 
\end{tabularx}
\setlength\extrarowheight{0pt}

\subsection{Tracer Parameters (group: \inlvar{/tracer/}, optional)}

\begin{multicols}{2}
\paragraph{Remarks:} The tracer parameter group contains information about the MPI tracers used during the experiment. For each of the $A$ tracers \inlvar{name}, \inlvar{batch}, \inlvar{vendor}, \inlvar{volume},  and molar \inlvar{concentration} of \inlvar{solute} per liter must be provided. Additionally, the time point of injection can be stored.

This version of the MDF can handle two basic scenarios. In the first one, static tracer phantoms are used. In this case, the phantom contains $A$ distinct tracers. For example, these might be particles of different core sizes, mobile or immobilized particles. In this case, \inlvar{injectionTime} is not used. In the second case, $A$ boli (e.g. pulsed boli) are administrated during the measurement, in which case the approximate administration volume, tracer type and time point of injection can be provided. Note that the injection clock recording the injection time should be synchronized with the clock, which provides the starting time of the measurement.

In case of a background measurement with no applied tracers in the scanner, the optional tracer group can be omitted.
\end{multicols}

\setlength\extrarowheight{5pt}
\noindent \begin{tabularx}{\columnwidth}{lllllX} 
\textbf{Parameter} & \textbf{Type} & \textbf{Dim} & \textbf{Unit/Format} & \textbf{Optional} & \textbf{Description} \\ \hline 
\inlvar{batch} & \inltab{String} & $A$ & & no & Batch of tracer \\ \hline
\inlvar{concentration} & \inltab{Float64} & $A$ & mol(\inlvar{solute})/L & no & Molar concentration of \inlvar{solute} per litre \\ \hline
\inlvar{injectionTime} & \inltab{String} & $A$ & yyyy-mm-ddThh:mm:ss.ms & yes & UTC time at which tracer injection started \\ \hline
\inlvar{name} & \inltab{String} & $A$ & & no & Name of tracer used in experiment \\ \hline
\inlvar{solute} & \inltab{String} & $A$ & & no & Solute, e.g. Fe \\ \hline
\inlvar{vendor} & \inltab{String} & $A$ & & no & Name of tracer supplier \\ \hline
\inlvar{volume} & \inltab{Float64} & $A$ & L & no & Total volume of applied tracer \\ \hline
\end{tabularx}
\setlength\extrarowheight{0pt}

\newpage
\subsection{Scanner Parameters (group: \inlvar{/scanner/}, non-optional)}

\begin{multicols}{2}
\paragraph{Remarks:} The scanner parameter group contains information about the MPI scanner used, such as \inlvar{name}, \inlvar{manufacturer}, \inlvar{boreSize}, field \inlvar{topology}, \inlvar{facility} where the scanner is installed, and the \inlvar{operator}.
\end{multicols}

\setlength\extrarowheight{5pt}
\noindent \begin{tabularx}{\columnwidth}{lllllX}
\noindent \textbf{Parameter} & \textbf{Type} & \textbf{Dim} & \textbf{Unit/Format} & \textbf{Optional} & \textbf{Description} \\ \hline 
\inlvar{boreSize} & \inltab{Float64} & 1 & m & yes & Diameter of the bore \\ \hline 
\inlvar{facility} & \inltab{String} & 1 & & no & Facility where the MPI scanner is installed \\ \hline 
\inlvar{manufacturer} & \inltab{String} & 1 & & no & Scanner manufacturer \\ \hline 
\inlvar{name} & \inltab{String} & 1 & & no & Scanner name \\ \hline 
\inlvar{operator} & \inltab{String} & 1 & & no & User who operates the MPI scanner \\ \hline 
\inlvar{topology} & \inltab{String} & 1 & & no & Scanner topology (e.g. FFP, FFL, MPS)\\ \hline 
\end{tabularx}
\setlength\extrarowheight{0pt}

\newpage
\subsection{Acquisition Parameters (group: \inlvar{/acquisition/}, non-optional)}

\begin{multicols}{2}
\paragraph{Remarks:} The acquisition parameter group can describe different imaging protocols and trajectory settings. The corresponding data is organized into general information, a subgroup containing information on the $D$ excitation channels, and a subgroup containing information on the $C$ receive channels.

The term \textit{frame} refers to the data collected during one acquisition period. Usually all data within a frame will be combined to reconstruct a single MPI image/tomogram. In the simplest scenario, this data is acquired during one \inlvar{drivefield/cycle}. If block-averaging is applied the amount of data captured stays the same while the acquisition time increases by a factor of \inlvar{numAverages}. In this document we will refer to the product of one \inlvar{drivefield/cycle} and \inlvar{numAverages} as \textit{one drive-field period}.

In case that the static selection field is changed over time one will measure several drive-field periods each with a different gradient field setting. In a stepped multi-patch sequence the gradient field will be spatially shifted and/or scaled in between drive-field periods and remain constant within. In such a scenario a full frame consists of \inlvar{numPeriodsPerFrame} ($J$) periods. The gradient scaling or shift within each period are described by the fields \inlvar{/acquisition/gradient} and \inlvar{/acquisition/offsetField}. The former has $J \times Y \times 3 \times 3$ entries whereas the later is of dimension $J \times Y \times 3$. While $J$ labels the drive-field periods, $Y$ can be used to describe dynamic multi-patch sequences where the gradient and offset field change during a drive-field period. Each period is discretized into $Y$ equidistantly spaced time intervals. The values in \inlvar{/acquisition/gradient} and \inlvar{/acquisition/offsetField} describe the gradient and offset field at the beginning of each time interval. In case $Y=1$ the fields at the beginning of the period $J$ are provided, which can be used to describe a static multi-patch sequence. Higher numbers of $Y$ allow the description of an arbitrarily fine grained gradient and offset field sequence.

It should be noted that the combined gradient and offset field can be expressed locally around the scanner center $\mathbf r_\text{c}\in \mathbb R^3$ via the Taylor expansion
\begin{equation}
	\mathbf H(\mathbf r) = \mathbf H(\mathbf r_\text{c}) + \mathbf J_{\mathbf H}(\mathbf r_\text{c}) (\mathbf r - \mathbf r_\text{c}) + \dots,
\end{equation}
where $\mathbf J_{\mathbf H}(\mathbf r_\text{c})$ is the Jacobian of $\mathbf H$ at $\mathbf r_\text{c}$. \inlvar{/acquisition/offsetField} stores the field $\mathbf H(\mathbf r_\text{c})$ and \inlvar{/acquisition/gradient} stores the $3\times 3$ matrix $\mathbf J_{\mathbf H}$. Both parameters are for most scanner topologies sufficient to uniquely describe the applied field sequence. In case of non-linear fields and/or a higher degree of freedom in the number of applied field generators, one may need additional custom fields in order to uniquely define the sequence parameters.

\end{multicols}

\setlength\extrarowheight{5pt}
\noindent \begin{tabularx}{\columnwidth}{lllllX}
\textbf{Parameter} & \textbf{Type} & \textbf{Dim} & \textbf{Unit/Format} & \textbf{Optional} & \textbf{Description} \\ \hline
\inlvar{gradient} & \inltab{Float64} & $J \times Y \times 3 \times 3$ & Tm$^{-1}\mu_0^{-1}$ & yes & Gradient strength of the selection field in $x$, $y$, and $z$ directions \\ \hline
\inlvar{numAverages} & \inltab{Int64} & 1 & 1& no & Number of block averages per drive-field period. \\ \hline
\inlvar{numFrames} & \inltab{Int64} & 1 & 1& no & Number of available measurement frames $N$ \\ \hline
\inlvar{numPeriodsPerFrame} & \inltab{Int64} & 1 & 1 & no & Number of drive-field periods within a frame denoted by $J$ \\ \hline
\inlvar{offsetField} & \inltab{Float64} & $J \times Y \times 3$ & T$\mu_0^{-1}$ & yes & Offset field applied  \\ \hline
\inlvar{startTime} & \inltab{String} & 1 & yyyy-mm-ddThh:mm:ss.ms & no & UTC start time of MPI measurement \\ \hline
\end{tabularx}
\setlength\extrarowheight{0pt}

\newpage

\subsubsection{Drive Field Parameters (group: \inlvar{/acquisition/drivefield/}, non-optional)}

\begin{multicols}{2}
\paragraph{Remarks:} The drive field subgroup describes the excitation details of the imaging protocol. On the lowest level, each MPI scanner contains $D$ channels for excitation. Since most drive-field parameters may change from period to period, they have a leading dimension $J$.

These excitation signals are usually sinusoidal and can be described by $D$ amplitudes (drive field strengths), $D$ phases, a base frequency, and $D$ dividers. In a more general setting, the generated drive field of channel $d$ can be described by
$$
H_d(t) = \sum_{l=1}^{F} A_l \Lambda_l (2\pi f_l t + \varphi_l)
$$
where $F$ is the number of frequencies on the channel, $A_l$ is the drive-field strength, $\phi_l$ is the phase, $f_l$ is the frequency (\inlvar{baseFrequency}/\inlvar{divider}$_l$), and $\Lambda_l$ is the waveform. The waveform is specified by a dedicated parameter \inlvar{waveform}. It can be set to \textit{sine}, \textit{triangle}, or \textit{custom}. 
\end{multicols}

\setlength\extrarowheight{5pt}
\noindent \begin{tabularx}{\columnwidth}{lllllX} 
\textbf{Parameter} & \textbf{Type} & \textbf{Dim} & \textbf{Unit/Format} & \textbf{Optional} & \textbf{Description} \\ \hline 
\inlvar{baseFrequency} & \inltab{Float64} & 1 & Hz & no & Base frequency to derive drive field frequencies \\ \hline
\inlvar{cycle} & \inltab{Float64} & 1 & s & no & Trajectory cycle is determined by lcm(\inlvar{divider})/\inlvar{baseFrequency}. It will not change when averaging was applied. The duration for measuring the $V$ data points (i.e. the drive-field period) is given by the product of  \inlvar{period} and \inlvar{numAverages} \\ \hline
\inlvar{divider} & \inltab{Int64} & $D \times F$ & 1 & no & Divider of the \inlvar{baseFrequency} to determine the drive field frequencies \\ \hline
\inlvar{numChannels} & \inltab{Int64} & 1 & 1 & no & Number of drive field channels, denoted by $D$ \\ \hline
\inlvar{phase} & \inltab{Float64} & $J \times D \times F$ & rad, $[-\pi,\pi)$ & no & Applied drive field phase $\varphi$\\ \hline
\inlvar{strength} & \inltab{Float64} & $J \times D \times F $ & T$\mu_0^{-1}$ & no & Applied drive field strength \\ \hline
\inlvar{waveform} & \inltab{String} & $D \times F$ & 1 & no & Waveform type: \textit{sine}, \textit{triangle} or \textit{custom} \\ \hline
\end{tabularx}
\setlength\extrarowheight{0pt}

\subsubsection{Receiver (group: \inlvar{/acquisition/receiver/}, non-optional)}

\begin{multicols}{2}
	\paragraph{Remarks:} The receiver subgroup describes the details on the MPI receiver. For a multi-patch sequence, it is assumed, that the signal acquisition only takes place during particle excitation. During each drive-field cycle, $C$ receive channels record some quantity related to the magnetization dynamic. In most cases these, will be continuous voltage signals induced into the $C$ receive coils, which are proportional to the change of the particle magnetization. This analog signal will usually be converted by some sort of analog to digital converter (ADC) to a discrete series of integer numbers $r_{ci}$ for each channel $c$. To map the these values to the MPI measurement signal $u_{ci}^{ADC}$, one has to scale the numbers $r_{ci}$ and add an offset factor
\begin{equation*}
	u_{ci}^{ADC} = a_c r_{ci} +b_c.
\end{equation*}
Here, $a_c$ and $b_c$ are the scaling and offset factors corresponding to channel $c$, which can be stored in \inlvar{dataConversionFactor}. In case the conversion was already performed the \inlvar{dataConversionFactor} can be ommited.

The MPI measurement signal is acquired at $V$ equidistant time points. For inductive measurement systems the signal is usually not measured directly at the receive coils but amplified and filtered first, which may damp and distort the signal. Therefore, a transfer function can be stored in the parameter \inlvar{transferFunction}, which relates the Fourier domain voltage induced at the receive coil $\hat{u}_k^\text{coil}$ and the Fourier domain voltage $\hat{u}_k^\text{ADC}$ measured at the ADC by
\begin{equation*}
\hat{u}_k^\text{ADC} = a_k  \hat{u}_k^\text{coil}, \quad k=1,\dots,K.
\end{equation*}
Here, $a_k$ are the unitless parameters stored in \inlvar{transferFunction} for each receive channel individually. 

For MPS systems one can additionally store a parameter that maps the induced voltage to the mean magnetic moment of a magnetic nanoparticle located at the center of the scanner. More precisely, in each receive coil a projection of the mean magnetic moment onto the coil sensitivity is measured. The relation of this projection and the voltage at the receive coil in frequency space representation is given by
\begin{align*}
\hat{u}_k^\text{coil} = 2\pi \textrm{i} k \beta \hat{m}_k^\text{proj} , \quad k=1,\dots,K.
\end{align*}
where  
$\hat m_k^\text{proj}$ is the orthogonal projection of the magnetic moment onto the coil sensitivity at the scanner center
and $\beta$ is the channel dependent conversion factor that is stored in the parameter  \inlvar{inductionFactor}.
\end{multicols}

\setlength\extrarowheight{5pt}
\noindent \begin{tabularx}{\columnwidth}{lllllX} 
\textbf{Parameter} & \textbf{Type} & \textbf{Dim} & \textbf{Unit/Format} & \textbf{Optional} & \textbf{Description} \\ \hline 
\inlvar{bandwidth} & \inltab{Float64} & $1$ & Hz & no & Bandwidth of the receiver unit \\ \hline
\inlvar{dataConversionFactor} & \inltab{Float64} & $C \times 2$ & \inlvar{unit} & yes & Dimension less scaling factor and offset $(a_c, b_c)$ to convert raw data into a physical quantity with corresponding unit of measurement \inlvar{unit} \\ \hline 
\inlvar{inductionFactor} & \inltab{Float64} & $C$ & \inlvar{unit} A$^{-1}$m$^{-2}$ & yes & Induction factor mapping the projection of the magnetic moment to the voltage in the receive coil.  \\ \hline
\inlvar{numChannels} & \inltab{Int64} & 1 & & no & Number of receive channels $C$ \\ \hline 
\inlvar{numSamplingPoints} & \inltab{Int64} & $1$ &  & no & Number of sampling points during one period, denoted by $V$ \\ \hline
\inlvar{transferFunction} & \inltab{Complex128} & $C \times K$ & & yes & Transfer function of the receive channels in Fourier domain. \inlvar{unit} is the field from the \inlvar{/measurement} group \\ \hline
\inlvar{unit} & \inltab{String} & $1$ & & no & SI unit of the measured quantity, usually Voltage V \\ \hline 
\end{tabularx}
\setlength\extrarowheight{0pt}

\subsection{Measurement (group: \inlvar{/measurement/}, optional)}
\begin{multicols}{2}

\paragraph{Remarks:}
MPI data is usually acquired by a series of foreground measurements and optional background measurements. Here, we refer to background measurements as MPI data captured, when any signal generating material, e.g. a phantom or a delta sample is removed from the scanner bore. Initially, all data is available in time domain, where the data of a single frame consists of the signal recorded for all periods in each receive channel, i.e. $J \times C \times V$ data points per set with the temporal index being the fastest to access.  If several measurements are acquired (indicated by \inlvar{numFrames}), the frame dimension is the slowest to access. Along this dimension, the frames are ordered with respect to the time at which they were acquired starting with the measurement acquired first and stopping with the measurement acquired last. We refer to this data as raw measurement data. In Fourier representation, each frame would be stored by $J \times C\times K$ complex data points and $K = V/2 +1$.

Often it is not convenient to store the raw data but to perform certain processing steps and store the processed data. These steps may lead to a reduction in the number of sampling points from $V$ to $W$ or a corresponding reduction of frequency components $K$ depending on the final representation in which the raw/processed data is stored. The most common processing steps are:
\begin{enumerate}
	\item Spectral leakage correction, which may be applied to ensure that each individual frame is periodic.
	\item Background correction, where the background signal in subtracted.
	\item Fourier transformation bringing the data from time into the Fourier representation and storing them in Fourier representation.
	\item Transfer function correction to obtain the magnetic moment or induced voltage that has been measured.
	\item Frequency selection to reduce the number of frequency components, e.g. bandwidth reduction or selection of high-signal frequency components.
	\item Dimension permutation, which is usually applied to Fourier transformed data exchanging the storing order of the data for fast access to the frames.
	\item Frame permutation to reorder the frames within the data set.
	\item Compression of the $O$ foreground frames by applying a sparsity transformation and storing only significant coefficients.
\end{enumerate}
For each of the steps above there is a corresponding flag within this group indicating if the corresponding processing step has been carried out. 

During processing one might want to keep track which of the final $N$ frames belong to background measurements and which do not. To this end, the binary mask \inlvar{isBackgroundFrame} should be used. If frequency selection has been performed, \inlvar{frequencySelection} stores the $K$ frequency components (subset) selected from the set of acquired frequency components. If performed, a frame permutation can be described by a bijective mapping $\sigma : \left\{ 1,2,\dots,N \right\} \rightarrow \left\{ 1,2,\dots,N \right\}$ from the set of frame indices to itself. If such a permutation is performed, $\sigma$ is stored in the one-line notation as $\sigma(1)$, $\sigma(2)$, $\dots$, $\sigma(N)$ in \inlvar{framePermutation}.

The motivation for reducing the number of frequency components is the reduction of the file size. Another even more effective method is to compress the data by applying a (linear) sparsity transformation to the data along the frame dimension and storing only the most significant of the resulting coefficients. This method was initially proposed in \cite{lampe2012fast} where a global compression of the system matrix was used. This specification follows the concept introduced in \cite{knopp2015compression}, where a fixed number of coefficients is stored for each frequency component. 
The compression option, indicated by the flag \inlvar{isSparsityTransformed}, is only available if the processing stages indicated by \inlvar{isFastFrameAxis}, \inlvar{isFourierTransformed} have been applied. Furthermore, the frames have to be ordered in such a way that the leading $O$ frames are the foreground frames and the last $E$ frames are the background frames. The order of the frames can be verified by checking the boolean vector \inlvar{isBackgroundFrame}. The sparsity transformation is only applied to the $O$ foreground frames. 

Let $\hat{u}_{j,c,k,o}$ be the measured foreground data in frequency domain with $j=1,\dots,J$, $c=1,\dots,C$, $k=1,\dots,K$, and $o=1,\dots,O$. The processing indicated by \inlvar{isSparsityTransformed} is mathematically described by 
\begin{equation*}
 \tilde{u}_{j,c,k,n} = \sum_{o=1}^{O} \Gamma_{n,o} \hat{u}_{j,c,k,o}, \quad \text{for} \quad n=1,\dots,O
\end{equation*}
where $\mathbf{\Gamma} = \left( \Gamma_{n,o} \right)_{n,o=1,\dots,O}$ is the transformation matrix.
Instead of storing $\tilde{u}_{j,c,k,n}$ completely, the field \inlvar{data} contains a subset $\tilde{u}_{j,c,k,\beta_{j,c,k,b}}$ where $\beta_{j,c,k,b}$ with $b=1,\dots,B$ contains the subsampling indices, which are stored in the field \inlvar{subsamplingIndices}. In order to recover the data $\hat{v}_{j,c,k,o} \approx \hat{u}_{j,c,k,o}$, one first needs to restore $\tilde{v}_{j,c,k,n}$ by 
\begin{equation*}
	\tilde{v}_{j,c,k,n} = 
	\begin{cases}
		\tilde{u}_{j,c,k,n}, & \text{if}\ n =\beta_{j,c,k,b} \\
		0, & \text{otherwise.}
	\end{cases}
\end{equation*}
Then, the inverse transformation $\mathbf{\Gamma}^{-1}$ needs to be applied. In case of unitary transformations, the inversion can be done by applying the adjoint transformation $\mathbf{\Gamma}^{\adj}$, i.e.
\begin{equation*}
 \hat{v}_{j,c,k,o} = \sum_{n=1}^{O} \overline{\Gamma}_{n,o} \tilde{v}_{j,c,k,n}, \quad \text{for} \quad o=1,\dots,O
\end{equation*}
We note that $\hat{v}_{j,c,k,o}$ is only an approximation of $\hat{u}_{j,c,k,o}$ and that the accuracy of this approximation depends on the compression rate that is given by $\frac{B}{O}$.

The name of the applied transformation $\mathbf{\Gamma}$ is stored in the field \inlvar{sparsityTransformation}. Currently, the following transformations are supported: DCT-I, DCT-II, DCT-III, DCT-IV. In all cases we consider the orthogonal / unitary variant of the transformation. For details on the precise definition of the transformations, we refer to the documentation of the FFTW library \cite{FFTW05}. Additional transformations are not standardized and must be integrated into this specification before use. In case that the $O$ foreground frames lay on a 2D or 3D grid we consider the multidimensional versions of the transformations. The dimensionality of the transformation can be derived from the number of non-singleton dimensions in \inlvar{/calibration/size}. 

\end{multicols}

\setlength\extrarowheight{5pt}
\noindent \begin{tabularx}{\columnwidth}{llp{3cm}lX} 
\textbf{Parameter} & \textbf{Type} & \textbf{Dim} &  \textbf{Optional} & \textbf{Description} \\ \hline 
\inlvar{data} & \inltab{Number} & $N \times J \times C \times K$ or $ J \times C \times K\times N$ or $N \times J \times C \times W$ or $ J \times C \times W \times N$ or $J \times C \times K \times (B+E)$ & no & Measured data at a specific processing stage \\ \hline
\inlvar{framePermutation} & \inltab{Int64} & $N$ & \inlvar{isFramePermutation} & Indices of original frame order\\ \hline
\inlvar{frequencySelection} & \inltab{Int64} & K & \inlvar{isFrequencySelection} & Indices of selected frequency components \\ \hline
\inlvar{isBackgroundCorrected} & \inltab{Int8} & 1 & no & Flag, if the background has been subtracted \\ \hline
\inlvar{isBackgroundFrame} & \inltab{Int8} & $N$ & no & Mask indicating for each of the $N$ frames if it is a background measurement (true) or not \\ \hline
\inlvar{isFastFrameAxis} & \inltab{Int8} & 1 & no & Flag, if the frame dimension $N$ has been moved to the last dimension\\ \hline
\inlvar{isFourierTransformed} & \inltab{Int8} & 1 & no & Flag, if the data is stored in frequency space \\ \hline
\inlvar{isFramePermutation} & \inltab{Int8} & 1 & no & Flag, if the order of frames has been changed, see \inlvar{framePermutation} \\ \hline 
\inlvar{isFrequencySelection} & \inltab{Int8} & 1 & no & Flag, if only a subset of frequencies has been selected and stored, see \inlvar{frequencySelection}\\ \hline 
\inlvar{isSparsityTransformed} & \inltab{Int8} & 1 & no & Flag, if the foreground frames are compressed along the frame dimension \\ \hline
\inlvar{isSpectralLeakageCorrected} & \inltab{Int8} & 1 & no & Flag, if spectral leakage correction has been applied \\ \hline
\inlvar{isTransferFunctionCorrected} & \inltab{Int8} & 1 & no & Flag, if the data has been corrected by the \inlvar{transferFunction}\\ \hline 
\inlvar{sparsityTransformation} & \inltab{String} & 1 & \inlvar{isSparsityTransformed} & Name of the applied sparsity transformation\\ \hline
\inlvar{subsamplingIndices} & \inltab{Integer} & $ J \times C \times K \times B$ & \inlvar{isSparsityTransformed} & Subsampling indices $\beta_{j,c,k,b}$\\ \hline
\end{tabularx} 
\setlength\extrarowheight{0pt}

\newpage
\subsection{Calibration (group: \inlvar{/calibration/}, optional)}

\begin{multicols}{2}
\paragraph{Remarks:}
The calibration group describes a calibration measurement (system matrix), although it does not hold the data itself. Each of the calibration measurements is taken with a calibration sample (delta sample) at a grid centered position inside the FOV of the device. If available, background measurements are taken with the delta sample outside of the FOV of the scanner. Usually, not the raw measurements are stored but processed data, where at least averaging, Fourier transformation, frame permutation and transposition has been performed yielding a total of $N$ processed frames. Of these frames $O$ frames correspond to the calibration scans at the $O$ spatial positions, whereas the other frames correspond to background measurements taken throughout the calibration process. All processing steps are documented in the \inlvar{/measurement} group.

If the measurements were taken on a regular grid of size $N_x \times N_y \times N_z$, the permutation is usually done such that measurements are ordered with respect to their $x$ position first, second with respect to their $y$ position, and last with respect to their $z$ position. Background measurements are collected at the end in \inlvar{/measurement/data}, which in combination with reordering of the measurements allows a fast access to the system matrix. If a different storage order is used this can be documented using the optional parameter \inlvar{order}. For non-regular sampling points, there is the possibility to explicitly store all $O$ positions. If the calibration measurement is performed in a MPS system, the spatial positions are usually emulated by applying offset fields, which can be stored in \inlvar{offsetFields}.
\end{multicols}

\setlength\extrarowheight{5pt}
\noindent \begin{tabularx}{\columnwidth}{llp{3cm}llX} 
\textbf{Parameter} & \textbf{Type} & \textbf{Dim} & \textbf{Unit/Format} & \textbf{Optional} & \textbf{Description} \\ \hline 
\inlvar{deltaSampleSize} & \inltab{Float64} & 3 & m & yes & Size of the delta sample used for calibration scan \\ \hline
\inlvar{fieldOfView} & \inltab{Float64} & $3$ & m & yes & Field of view of the system matrix \\ \hline
\inlvar{fieldOfViewCenter} & \inltab{Float64} & $3$ & m & yes & Center of the system matrix (relative to origin/center) \\ \hline
\inlvar{method} & \inltab{String} & 1 & & no & Method used to obtain calibration data. Can for instance be robot, hybrid, or simulation \\ \hline
\inlvar{offsetFields} & \inltab{Float64} & $O \times 3$ & T$\mu_0^{-1}$ & yes & Applied offset field strength to emulate a spatial position \mbox{($x$, $y$, $z$)}\\ \hline
\inlvar{order} & \inltab{String} & 1 & & yes & Ordering of the dimensions, default is \textit{xyz} \\ \hline
\inlvar{positions} & \inltab{Float64} & $O \times 3$ & m & yes & Position of each of the grid points, stored as \mbox{($x$, $y$, $z$)} triples \\ \hline
\inlvar{size} & \inltab{Int64} & $3$ &  & yes & Number of voxels in each dimension, inner product is $O$ \\ \hline
\inlvar{snr} & \inltab{Float64} & $J \times C \times K$ &  & yes & Signal-to-noise estimate for recorded frequency components \\ \hline
\end{tabularx}
\setlength\extrarowheight{0pt}

\newpage
\subsection{Reconstruction Results (group: \inlvar{/reconstruction/}, optional)}

\begin{multicols}{2}
Reconstruction results are stored using the parameter \inlvar{data} inside this group. \inlvar{data} contains a $Q\times P \times S$ array, where $Q$ denotes the number of reconstructed frames within the data set, $P$ denotes the number of voxels and $S$ the number of multispectral channels. If no multispectral reconstruction is performed, then one may set $S=1$. Depending on the reconstruction the grid of the reconstruction data can be different from the system matrix grid. Hence, grid parameters are mirrored in the \inlvar{/reconstruction} group.

For analysis of the MPI tomograms, it is often required to know which parts of the reconstructed tomogram have been covered by the trajectory of the field free region. In MPI, one refers to the non-covered region as overscan region. Therefore, the optional binary field \inlvar{isOverscanRegion} stores for each voxel if it is part of the overscan region. If no voxel lies within the overscan region, \inlvar{isOverscanRegion} may be omitted.
\end{multicols}

\setlength\extrarowheight{5pt}
\noindent \begin{tabularx}{\columnwidth}{lllllX} 
\textbf{Parameter} & \textbf{Type} & \textbf{Dim} & \textbf{Unit/Format} & \textbf{Optional} & \textbf{Description} \\ \hline 
\inlvar{data} & \inltab{Number} & $Q\times P \times S$ & & no & Reconstructed data \\ \hline
\inlvar{fieldOfView} & \inltab{Float64} & $3$ & m & yes & Field of view of reconstructed data \\ \hline
\inlvar{fieldOfViewCenter} & \inltab{Float64} & $3$ & m & yes & Center of the reconstructed data (relative to scanner origin/center) \\ \hline 
\inlvar{isOverscanRegion} & \inltab{Int8} & $P$ &  & yes & Mask indicating for each of the $P$ voxels if it is part of the overscan region (true) or not \\ \hline
\inlvar{order} & \inltab{String} & 1 & & yes & Ordering of the dimensions, default is \textit{xyz} \\ \hline
\inlvar{positions} & \inltab{Float64} & $P \times 3$ & m & yes & Position of each of the grid points, stored as ($x$, $y$, $z$) tripels \\ \hline
\inlvar{size} & \inltab{Int64} & $3$ &  & yes & Number of voxels in each dimension, inner product is $P$ \\ \hline
\end{tabularx}
\setlength\extrarowheight{0pt}

\clearpage
\section{Changelog}

\begin{multicols}{2}
\subsection{v2.1.0}

\begin{itemize}
	\item Added the possibility to store compressed calibration data using a sparsity transformation. To this end, the \inlvar{data} field in the  \inlvar{measurement} group has been extended and the fields \inlvar{isSparsityTransformed}, \inlvar{sparsityTransformation} and \inlvar{subsamplingIndices} have been introduced.
\end{itemize}

\subsection{v2.0.1}

\begin{itemize}
	\item Added optional \inlvar{time} field to \inlvar{study} group to be able to document the creation time of the study.
	\item Lexicographical reordering of all fields within each group to improve readability.
\end{itemize}

\subsection{v2.0.0}

\begin{itemize}
	\item Version 2 of the MDF is a major update breaking backwards compatibility with v1.x. The major update was necessary due to several shortcomings in the v1.x.
	\item The naming of parameters was made more consistent. Furthermore, some parameters moved from one group into another.
	\item Defined a complex datatype using a HDF5 compound type. 
	\item In v1.x it was not possible to store background data. This functionality has been added in v2.
	\item We simplified the measurement group and made it much more expressive. In v1.x it was not entirely clear, which processing steps have been applied to the measurement data in the stored dataset. The \inlvar{measurement} group now contains several flags that precisely document the state of the stored data. Using this it is now possible to also store calibration data in the \inlvar{measurement} group. The calibration group in turn only stores metadata about a calibration experiment while the actual data is store in the measurement group.
	\item Updated Affiliations in the MDF specification.
	\item Improved the general descriptions of fields and groups.
	\item In v1.x the MDF allowed many fields to have varying dimensions depending on the context. As of version 2.0.0 only one field offers this freedom. This change should make implementations handling MDF files less complex.
	\item \inlvar{Number} has been introduced as a generic type. 
	\item Added a table listing all variable names used in the descriptions of parameters.
	\item Added a section describing the possibility to add user defined parameters to MDF files.
	\item Added a description for optional and non-optional groups and conditional, optional, and non-optional datasets.
    \item Added a short section on the code examples on the Github repository.
	\item Support for triangle wave forms has been added.
	\item Support for multiple excitation frequencies on a drive-field channel has been added.
	\item Added the dimension $A$ to all fields of the \inlvar{tracer} group to be able to describe settings where multiple tracers are used or tracers are administered multiple times.
	\item Added the possibility to store the tracer concentration also for non iron based tracer materials by adding the \inlvar{/tracer/solute} field and redefining the field \inlvar{tracer/concentration}.
 	\item Improved documentation for the storage of ADC transfer functions. It is also possible now to store the measurement data as integer data and use a \inlvar{dataConversionFactor} to describe the mapping to a physical representation (e.g. Volt)
 	\item Added support for receive coil to ADC transfer functions.
 	\item Added support for mean magnetic moment to receive coil voltage transfer functions.
	\item Split the \inlvar{/study} group into the \inlvar{/experiment} group and the study group. This allows to provide more fine grained information on study and experiment.
	\item Added possibility to mark the overscan region.
	\item Added new section changelog to the MDF documentation to record the development of the MDF.
	\item Updated \inlvar{README.md} and \inlvar{MDF.bib} in the github repository.
	\item Updated code examples in the github repository.
	\item Added a section on the MDF reference implementation MPIFiles.jl. Since sanity checks will be covered by this package, the description on sanity checks has been removed.
\end{itemize}

\subsection{v1.0.5}

\begin{itemize}
	\item Added the possibility to store different channels of reconstructed data.
	\item Added support for receive channels with different characteristics (e.g. bandwidth).
	\item Made dataset \inlvar{/acquisition/receiver/frequencies} optional.
	\item Extended the description on the data types, which are used to store data.
	\item Added references for Julia and HDF5 to the specifications.
\end{itemize}

\subsection{v1.0.4}

\begin{itemize}
	\item Clarify that HDF5 datasets are used to store MPI parameters.
\end{itemize}

\subsection{v1.0.3}

\begin{itemize}
	\item Updated Affiliations in the MDF specification.
	\item Included data download into the Python and Matlab example code.
	\item Changes in the Python and Matlab example code to be better comparable to the Julia example code.
\end{itemize}

\subsection{v1.0.2}

\begin{itemize}
	\item Added reference to arXiv paper and bibtex file for reference.
\end{itemize}

\subsection{v1.0.1}

\begin{itemize}
	\item A sanity check within the Julia code shipped alongside the specifications.
	\item An update to the specification documenting the availability of a sanity check.
	\item Updated MDF files on https://www.tuhh.de/ibi/research/mpi-data-format.html.
	\item Updated documentation to the Julia, Matlab and Python reconstruction scripts.
	\item Improved Julia reconstruction script, automatically downloading the required MDF files.
\end{itemize}
\end{multicols}

\newpage
\bibliographystyle{unsrt}
\bibliography{MDF}

\end{document}